\documentstyle[12pt,epsfig]{report}
\def\thebibliography#1{\leftline{\large\bf References}\list
  {[\arabic{enumi}]}{\settowidth\labelwidth{[#1]}\leftmargin\labelwidth
    \advance\leftmargin\labelsep
    \usecounter{enumi}}
    \def\newblock{\hskip .11em plus .33em minus .07em}
    \sloppy\clubpenalty4000\widowpenalty4000}

\newcommand{\be}{\begin{equation}}
\newcommand{\bea}{\begin{eqnarray}}
\newcommand{\ba}{\begin{array}}
\newcommand{\ea}{\end{array}}
\newcommand{\en}{\end{equation}}
\newcommand{\eea}{\end{eqnarray}}
\newcommand{\half}{\fract{1}{2}}

\newcommand{\dslash}{\partial \hskip -0.55em /}
\newcommand{\fract}[2]{{\textstyle\frac{#1}{#2}}} 

\begin{document}

\baselineskip24pt

\begin{center}{\Large \bf Searching for Quantum Solitons in a 3+1
Dimensional Chiral Yukawa Model}
\end{center}

\baselineskip16pt

\vskip1cm

\centerline{
E.~Farhi$^{\rm a,}$\footnote{e-mail:  farhi@mit.edu,
graham@physics.ucla.edu, jaffe@mit.edu, herbert.weigel@uni-tuebingen.de},
N.~Graham$^{\rm b}$,
R.L.~Jaffe$^{\rm a}$, and
H.~Weigel$^{\rm c,}$\footnote{Heisenberg Fellow}}

\parbox[t]{14cm}{
\begin{center}
{~}\\$^{\rm a}$Center for Theoretical Physics,\\
Laboratory for Nuclear Science and Department of Physics \\
Massachusetts Institute of Technology, Cambridge, Massachusetts 02139 \\~\\
$^{\rm b}$Department of Physics and Astronomy \\
University of California at Los Angeles \\
Los Angeles, CA  90095 \\~\\
$^{\rm c}$Institute for Theoretical Physics, T\"ubingen University\\
D-72076 T\"ubingen, Germany\\
{~} \\
{\rm MIT-CTP\#3223 \qquad UNITU-HEP-33/2001 \qquad UCLA/01/TEP/39 \qquad
hep-th/0112217}
\end{center}
}

\vskip1cm

\centerline{\bf ABSTRACT}
\vskip 0.2cm

\parbox[t]{15cm}{\small
We search for static solitons stabilized by heavy
fermions in a 3+1 dimensional Yukawa model.  We compute the
renormalized energy functional, including the exact one-loop quantum
corrections, and perform a variational search for configurations
that minimize the energy for a fixed fermion number.  We compute the
quantum corrections using a phase shift parameterization, in which we
renormalize by identifying orders of the Born series with
corresponding Feynman diagrams.  For higher-order terms in the Born
series, we develop a simplified calculational method.  When
applicable, we use the derivative expansion to check our results. We
observe marginally bound configurations at large Yukawa coupling, and
discuss their interpretation as soliton solutions subject to
general limitations of the model.}

\bigskip
\bigskip

\leftline{\small PACS: 05.45.Yv, 11.27.+d, 11.30.Rd, 11.10.Gh.}

\bigskip

\leftline{\small
Keywords:~\parbox[t]{11cm}{Solitons, Quantum Corrections,
Vacuum Polarization Energy, Renormalization, Variational Methods.}}

\newpage

\bigskip
\stepcounter{section}
\leftline{\large\bf 1. Introduction and Motivation}
\bigskip

Chiral gauge theories, such as the electroweak Standard Model, present a
challenge to conventional notions of decoupling.  By increasing its
Yukawa coupling, one can make a fermion so heavy that it should be
irrelevant to low-energy physics.  On the other hand, it cannot
simply disappear from the theory, since then anomalies would no longer
cancel. It is known that decoupling a chiral fermion leaves behind a
Wess-Zumino-Witten functional of the Higgs and gauge fields, which
keeps the path integral gauge invariant~\cite{decoupling}. However, to
cancel Witten's non-perturbative $SU(2)$ anomaly~\cite{Wi82}, an even
number of chiral doublets must be present in the low-energy
theory. After decoupling a fundamental fermion, the presence of a
fermionic soliton in the low-energy theory would ensure that gauge
invariance is maintained at the level of the
states~\cite{decoupling}.

There is a natural mechanism for realizing this scenario.  A twisted
configuration of the Higgs field will cause one fermion level to
become very tightly bound, or even to cross zero
energy~\cite{Ka84}. In the former case, the level can be filled at
very little cost in  energy, while in the latter case the background
field itself carries fermion number.  This energy must be added to the
classical energy required to twist the Higgs field, and then compared
to the mass of a fermion in an unperturbed background. However, one
must also include the contribution of the shift in the zero-point
energies of all the fermion modes, since it is of the same order in
$\hbar$ as the energy of the filled level.  If the total energy of a
fermion number one twisted configuration is below the mass of a free
fermion, then the configuration is stable.

In this paper, we explore this phenomenon in a simplified version
of the electroweak sector of the Standard Model.  We consider a Higgs
doublet chirally coupled to a single heavy fermion doublet.  We assume
the fermions in the doublet have equal masses, and consider a
hedgehog configuration for the Higgs field.  We ignore the
SU(2)$_{L}$-gauge fields. As we will argue in our conclusions, this
may be an important omission, and work is currently underway to extend
the calculation to include that case.

Previous work \cite{us1} showed that quantum-stabilized chiral
solitons do exist in the one-dimensional analog of the model we are
considering.  Here, as in that work, we will consider only the quantum
correction from the fermion loop, which we expect to be the most
important effect.  Neglecting bosonic loops is rigorously justified in
the large $N$ limit, where $N$ is the number of independent fermion
species, but here we couple to only a single doublet and take~$N=1$.

Our methods allow us to evaluate the one-loop fermion
contribution to the energy exactly, maintaining a fixed
renormalization scheme while exploring different Higgs
backgrounds. There have been earlier attempts to compute the one
fermion loop energy for a three-dimensional chiral background within
various approximation schemes.  Examples include discretization
methods~\cite{Al96,Ri87}, which may not be rigorously valid in the
continuum limit and require cutoffs, making the renormalization
obscure.  Other approaches use expansions that are valid for
slowly~\cite{Ai85} or rapidly~\cite{Ri87,Ba86} varying background
fields. Also truncations~\cite{Ad91} in the potential
generated by the background field, the heat kernel
expansion~\cite{Ka92}, and subsets of field configurations with less
severe ultraviolet divergences~\cite{Baa} have been studied.  While
these methods may be appropriate for specific applications and
useful within certain regions in configuration or parameter space,
they do not allow one  to explore a full range of {\it ans\"atze} for
the Higgs background or to make definitive statements about the
existence of a soliton.

We include in this Introduction a brief review of our method for
computing the contribution to the energy from vacuum polarization
induced by the background field.  For a fuller discussion of the
method see Ref.~\cite{Fa98,method1}.

The vacuum polarization energy is given formally by a sum over bound
states and an integral over continuum energies in the background given
by $\Phi$,
\begin{equation}
\frac{1}{2}\sum_{j}\epsilon_{j} +
\frac{1}{2} \int dk \frac{dn}{dk} \omega(k)
\end{equation}
where $\{\epsilon_{j}\}$ are the bound state energies,
$\omega(k)=\sqrt{k^{2}+m^{2}}$ is the energy of a scattering state,
and ${dn}/{dk}$ is the change the continuum density of states due to
the background field.  ${dn}/{dk}$ can be calculated from the
derivative of the phase shifts with respect to $k$~\cite{Schw},
\begin{equation}
    \frac{dn}{dk}=\frac{1}{2\pi i}\,{\rm tr}
 \ln S(k)=\frac{1}{\pi}\frac{d}{dk}\sum_{\ell}D(\ell)\delta(k,\ell) 
    \label{0.1}
\end{equation}
where we have expanded the $S$ matrix in partial waves
labeled by $\ell$, representing quantum numbers like angular
momentum and parity.  $\delta(k,\ell)$ gives the phase shift and
$D(\ell)$ the degeneracy.  Since we will consider cases where the
spectrum is asymmetric in energy, we define $\delta(k,l)$ to be the
sum of the phase shifts over both signs of the energy:
$\delta(k,l)= \delta_+(k,l) +\delta_-(k,l)$, where $\pm$ specifies
$\omega=\pm\sqrt{k^{2}+m^{2}}$.

Before we can do the $k$ integral, we must deal with potential
divergences.  We regulate the theory by dimensional regularization,
that is, we analytically continue the entire theory to $n$ dimensions,
where the integrals converge. For integer $n$, the expected divergences
emerge from the high momentum behavior of the phase shift integral.
At high momentum, the Born series becomes a good approximation to the
phase shift, so by subtracting successive terms in the Born
series from $\delta(k,l)$, we can remove terms from the
$k$-integration that would diverge at physical values of $n$.  The
expansion in the Born series can then be unambiguously identified with
the expansion of the effective energy in terms of Feynman diagrams
with insertions of the background field~\cite{Baa,Fa98,method1}.  In
noninteger dimensions, the contributions of both the Born terms and
the Feynman diagrams to the vacuum polarization energy are finite and
unambiguous analytic functions of $n$. Therefore, when we subtract a
term in the Born expansion and add back the equivalent Feynman
diagram, we can be certain that we are not introducing finite
ambiguities into the computation. The subtracted integration over
the density of states is then finite and we can take the limit to integer
$n$ without difficulty. Finally, we introduce the contributions from
the counterterms, which have been computed using standard
renormalization conditions in the perturbative sector of the model.
As usual, the potentially divergent pieces of the Feynman diagram are
canceled by the counterterm contributions. In all, the renormalized
vacuum polarization energy is given by
\begin{eqnarray}
    E_{\rm vac} &=& \pm \left(\sum_j D(j) \epsilon_j
    + \int_0^\infty \frac{dk}{2\pi}\sqrt{k^2+m^2}\frac{d}{dk}
    \sum_l D(l)\left(\delta(k,l)-\sum_{N=1}^{N_{\rm max}}
    \delta^{(N)}(k,l)\right) \right)
    \nonumber \\*&&
   \qquad{} +\sum_{N=1}^{N_{\rm max}} \Gamma^{(N)}(\Lambda)
    +\Gamma_{\rm ct}(\Lambda) 
    \label{ecasgeneral}
\end{eqnarray}
for bosons and fermions respectively, where the $N^{\rm th}$-order
Born approximant to the phase shifts is denoted by
$\delta^{(N)}(k,l)$.  $N_{\rm max}$ is the number of Born subtractions
required to render the $k$ integration finite.  The compensating
Feynman diagrams are denoted by $\Gamma^{(N)}(\Lambda)$, and
$\Gamma_{\rm ct}(\Lambda)$ represents the contribution of the
counterterms.  Both are cutoff-dependent, but as usual,
renormalization conditions will determine an unambiguous, finite
result.  We are left with two finite and numerically tractable
objects, the momentum integral and the sum $\sum_{N=1}^{N_{\rm max}}
\Gamma^{(N)}+\Gamma_{\rm ct}$.  As a result, no explicit cutoff
needs to be introduced in the numerical computation.

This paper is organized as follows. In Section 2 we briefly review the
Higgs sector of the standard model and outline its connection to the
Yukawa model. In Section 3 we discuss the renormalization of the
fermion loop and describe our calculation of the associated
contribution to the energy. Section 4 contains the numerical
analysis. We summarize and provide an outlook on future studies in
Section 5. Technical details are given in three Appendices. In
Appendix A we explore the Dirac equation and its scattering solutions
for a chiral background field. In Appendix B we describe and
numerically verify a simplified treatment of contributions
to the renormalized vacuum polarization energy  that are higher order
in the background field. In Appendix C we derive results in the
derivative expansion, which we use to check our results in the case of
slowly varying background fields.

\bigskip
\bigskip
\stepcounter{section}
\leftline{\large\bf 2. The Model}
\bigskip

The model we consider consists of the Higgs sector of the
Standard Model coupled to a fermion doublet in $3+1$ dimensions. Our
goal is to explore the possibility that within this model
there is a non-trivial Higgs field configuration with nonzero fermion
number whose energy is less than that of a state with the same quantum
numbers built on top of the perturbative vacuum.  The fermions get
their masses through their Yukawa coupling
to the Higgs.  Our model differs in two essential ways from the
Standard Model: we omit gauge fields and our fermions have equal
masses.  At the end of the paper we discuss the possible sensitivity
of our results to the omission of gauge fields.

We write the Higgs sector of the Standard Model in terms of a
$2\times2$ matrix-valued Higgs field
\begin{equation}
    \Phi=\pmatrix{\varphi_0 & -\varphi_+^* \cr
    \varphi_+& \varphi_0^*}\,
    \label{defhiggs}
\end{equation}
where $(\varphi_0, \varphi_+)$ is the usual doublet.  The Higgs
Lagrangian is
\begin{equation}
    {\cal L}_H=\frac{1}{4}\, {\rm tr}\left[\partial_\mu\Phi^\dagger
    \partial^\mu\Phi\right] -V(\Phi)
    \label{lhiggs}
\end{equation}
where
\begin{equation}
    V(\varphi)=\frac{\lambda}{16}\left(
    {\rm tr}\left[\Phi^\dagger\Phi\right]-2v^2\right)^2
    \label{higgspot}\ .
\end{equation}
We take the vacuum expectation value to be
\begin{equation}
    \langle\Phi\rangle = v \pmatrix{1 & 0\cr 0 &1}
    \label{vachiggs}
\end{equation}
and note that the Higgs particle has mass $m_H=\sqrt{\lambda} v$.

The coupling to the fermion doublet $q=(t,b)$ is given by
\begin{equation}
    {\cal L}_{\rm HF}=g\bar{q}_L\Phi q_R+g \bar{q}_R\Phi^\dagger q_L
    \label{lhf1}
\end{equation}
which results in mass $m=gv$ for both $t$ and $b$. It is also
convenient to rewrite $\Phi$ in terms of four real (dimensionless)
fields $s$ and $\vec{p}$ as
\begin{equation}
    \Phi =v\left(s+i\vec{\tau}\cdot\vec{p}\right)
    \label{defsandp}
\end{equation}
which gives
\begin{equation}
    {\cal L}_{\rm
    HF}=m\bar{q}\left(s+i\gamma_5\vec{\tau}\cdot\vec{p}\right)q\ .
    \label{lhf2}
\end{equation}
With these definitions, the classical energy is
\begin{equation}
    E_{\rm cl}[\Phi]=\frac{v^2}{2}\int d^3r \left(
    \partial_i s\partial_i s
    +\partial_i \vec{p}\cdot\partial_i \vec{p}
    +\frac{\lambda v^2}{4}(s^2+\vec{p}\,^2-1)^2\right)\ .
    \label{ecl1}
\end{equation}

\bigskip
\bigskip
\stepcounter{section}
\leftline{\large\bf 3. The Fermion Loop}
\bigskip

In this section we discuss the contribution of the fermion vacuum to
the total energy. This contribution arises because the fermionic
vacuum is polarized by the Higgs background.  In order to compute this
contribution we first have to outline the renormalization process in
the perturbative sector of the model.  The divergences of our model
can be canceled by counterterms of the form
\begin{equation}
    {\cal L}_{\rm ct}=a\,{\rm tr}\,
    \left(\partial_\mu\Phi\partial^\mu\Phi^\dagger\right)
    -b\,{\rm tr}\,\left(\Phi\Phi^\dagger-v^2\right)
    -c\,{\rm tr}\,\left(\Phi\Phi^\dagger-v^2\right)^2 
    \label{lct}
\end{equation}
where $a$, $b$, and $c$ are cutoff-dependent constants.  The Yukawa
coupling $g$, and consequently the fermion mass $m$, are not
renormalized at this order.

In terms of the shifted Higgs field $h \equiv s-v$,
our renormalization conditions are that the vacuum
expectation value of $h$ vanishes, and that the fermion loop
changes neither the position $m_H$ nor the residue of the pole
in the two-point function for $h$.  In order to fix the counterterms,
it is therefore sufficient to expand\footnote{Here and in what follows
${\rm tr}$ refers to sums over discrete labels while ${\rm Tr}$
includes the space-time integration.}
\begin{equation}
    {\cal S}_{\rm eff}[h]=-i\,{\rm Tr}\, \ln \, \left\{i\dslash
    - g (v+h)\right\}
    \label{floop}
\end{equation}
up to quadratic order in $h$ and combine the result with $\int d^4x
{\cal L}_{\rm ct}$. In dimensional regularization we obtain
\begin{eqnarray}
    a&=&-\frac{g^2}{(4\pi)^2}\left\{
    \frac{1}{\epsilon}-\gamma-\frac{2}{3}+
    \ln \,\left(\frac{4\pi\mu^2}{m^2}\right)
    -6\int_0^1dx x(1-x)\,\ln \,\left[1-x(1-x)
    \frac{m_H^2}{m^2}\right]\right\}
    \nonumber \\
    b&=&-\frac{g^2m^2}{(4\pi)^2}\left\{
    \frac{1}{\epsilon}-\gamma+1+\ln \,
    \left(\frac{4\pi\mu^2}{m^2}\right)\right\}
    \nonumber \\
    c&=&-\frac{g^4}{(4\pi)^2}\left\{
    \frac{1}{\epsilon}-\gamma+\ln \,
    \left(\frac{4\pi\mu^2}{m^2}\right)
    -\frac{m_H^2}{4m^2}
    -\frac{3}{2}\int_0^1 dx \, \ln \,\left[1-x(1-x)
    \frac{m_H^2}{m^2}\right]\right\}
    \label{lctcoef}
\end{eqnarray}
where $d=4-2\epsilon$, $\mu$ is the scale required to keep $g$
dimensionless.

Having set up the model in the perturbative sector, we now turn to
non-trivial field configurations.  We restrict our attention to the
spherical {\it ansatz\/} for the Higgs field,
\begin{equation}
    \Phi(\vec{x})=v\left[s(r)+i\vec{\tau}\cdot\hat{x}\, p(r)\right]
    \label{background}
\end{equation}
with $r=\sqrt{\vec{x}\,^2}$. With the standard form of
the Dirac matrices, the corresponding Dirac operator becomes
\begin{equation}
    h_D=\pmatrix{m s(r) & -i \vec{\sigma}\cdot\vec{\nabla}+
    m i\vec{\tau}\cdot\hat{x}\, p(r) \cr
    -i \vec{\sigma}\cdot\vec{\nabla}
    -m i\vec{\tau}\cdot\hat{x}\, p(r) & -m s(r)}
    \label{dirham}
\end{equation}
and the fermion field obeys the time-independent Dirac equation,
\begin{equation}
    h_D\Psi=\omega\Psi\  .
    \label{direq}
\end{equation}
Note that the energy eigenvalue $\omega$ can assume both positive
and negative values.  In general the spectrum of $h_{D}$ contains
discrete (bound) and continuum (scattering) states.

First we obtain the bound states $\epsilon_j$, the
solutions to eq.~(\ref{direq}) with $|\omega|<m$.  The numerical
method is sketched in Appendix A.  We can use Levinson's theorem to
compute the number of bound states in each channel from the phase
shifts.  The phase shifts are computed from the $S$-matrix,
which in turn is extracted from solutions to second-order
differential equations obtained from the Dirac equation.  Because we
restrict our attention to backgrounds in the spherical {\it ansatz\/},
there are two conserved quantum numbers, grand spin and parity.  The
grand spin $\vec G$ is defined as the vector sum of isospin and total angular
momentum (orbital plus spin), and can be interpreted as a generalized
angular momentum. The parity $\Pi$ is associated with space reflection
in the usual way.

We obtain second-order differential equations for the upper and lower
components of the Dirac equation in the standard basis. After
projecting onto a subspace with definite energy, grand spin and
parity, we have two coupled second-order differential equations for two
radial functions, $g_1(r)$ and $g_2(r)$.  Together, the linearly independent
solutions with incoming spherical waves in either of these channels
define a two-channel scattering problem.  In the following we will
suppress the labels $\omega$, $G$, and $\Pi$, which characterize this
two-dimensional problem.  The two linearly independent scattering
boundary conditions are labeled by $\{i,j\}=1,2$ and are implemented as
follows:  At large $r$ the solution $g_i^{(j)}$ has an outgoing wave if
$i=j$, and the radial wavefunction $i\ne j$ vanishes.  We summarize the two
wavefunctions and two boundary conditions in matrix form, ${\cal
G}_{ij}(r)=g_i^{(j)}(r)$.  We then write ${\cal G}(r)$ as a
multiplicative modification of the matrix solution to the free
differential equations,
\begin{equation}
    {\cal G}(r)=\pmatrix{g_1^{(1)}(r) & g_1^{(2)}(r)\cr
    g_2^{(1)}(r) & g_2^{(2)}(r)}\equiv
    F(r) H(kr) 
    \label{defF}
\end{equation}
where $H$ is diagonal and can be expressed simply in terms of Hankel
functions,
\begin{equation}
    H_+(x) = \pmatrix{h_G^{(1)}(x) & 0\cr
    0 & h_G^{(1)}(x)} \quad {\rm and}\quad
    H_-(x) = \pmatrix{h_{G+1}^{(1)}(x) & 0\cr
    0 & h_{G-1}^{(1)}(x)}
    \label{hankel}
\end{equation}
for $\Pi=+(-1)^G$ and $\Pi=-(-1)^G$, respectively.  For each value of grand
spin, $G$, the parity quantum number dictates the values of $\ell$ and
$\ell^{\prime}$ which enter $H_\pm={\rm
diag}(h_\ell^{(1)},h_{\ell^\prime}^{(1)})$.  In the channel with
parity $\Pi=+(-)^G$ we have $\ell=\ell^\prime=G$ while for
$\Pi=-(-1)^G$ we have $\ell=G+1$ and $\ell^\prime=G-1$.  Thus we have
The elements of the $2\times2$ matrix $F(r)$ satisfy second-order
differential equations obtained from the Dirac equation. They are of
the general form
\begin{equation}
    F^{\prime\prime}=-\frac{2}{r}F^\prime
    \left(1+rL^\prime(kr)\right)
    +\frac{s^\prime}{s\pm\omega/m}
    \left(F^\prime+F L^\prime(kr)\right)
    -V F +\frac{1}{r^2}\left[K,F\right] 
    \label{Fdifffeq}
\end{equation}
for upper and lower components respectively, where
\begin{equation}
   K=\pmatrix{\ell(\ell+1) & 0 \cr 0 & \ell^\prime(\ell^\prime+1)},
   \label{defK}
\end{equation}
with $\ell$ and $\ell^{\prime}$ as above.  $V$ is the $2\times2$
matrix describing the coupling of the fermions to the Higgs
background.  The particular forms of V are listed in Appendix A.  The
matrix $L=\ln \, H$ is the only remnant of the Hankel functions,
\begin{equation}
    L_+(x)=
    \pmatrix{\ln \, h_G^{(1)}(x) & 0 \cr
    0 &\ln \, h_G^{(1)}(x)} \quad {\rm and} \quad
    L_-(x)=
    \pmatrix{\ln \, h_{G+1}^{(1)}(x) & 0 \cr
    0 &\ln \, h_{G-1}^{(1)}(x)} \ .
    \label{defL}
\end{equation}
The elements of $L^\prime(x)=dL(x)/dx$ can be expressed as simple
rational functions, which avoids any instability in the numerical
treatment that would be caused by the oscillating Hankel functions.

The $2\times 2$ submatrix of the S-matrix can be constructed by superimposing
solutions to eq.~(\ref{Fdifffeq}).  First we normalize $F$ by
imposing the boundary conditions $F(r\to\infty)=1$ and
$F^\prime(r\to\infty)=0$.  Given these boundary conditions, since the
second-order differential equations for the $g_i$ are real, the
scattering wavefunction can be written as
\begin{equation}
    \Psi_{\rm sc}=-F^*(r)H^*(kr)+F(r)H(kr)S
    \label{scatwf}
\end{equation}
where $S$ is the $2\times 2$ submatrix of the $S$-matrix that we
are seeking.  Requiring that the scattering solution be regular
at the origin yields
\begin{equation}
    S=\lim_{r\to0}H^{-1}(kr)F^{-1}(r)F^*(r)H^*(kr)\ .
    \label{smatrix}
\end{equation}

The quantity that enters the density of states is the total phase shift
\begin{equation}
    \delta(k)=\frac{1}{2i}\, {\rm tr}\, \ln \, S
    = \frac{1}{2i} \lim_{r\to0}\, {\rm tr}\, \ln \,
    \left(F^{-1}(r)F^*(r)\right)
    \label{delta}
\end{equation}
from which $H$ cancels because as $r\to0$ the leading (singular) piece
of $H(kr)$ is real, {\it i.e.} $\lim_{r\to0}\,H^*(kr)H^{-1}(kr)=1$.
The unitarity of $S$ guarantees that equation~(\ref{delta}) explicitly
yields a real phase shift.

Eq.~(\ref{delta}) only gives the phase shift modulo $\pi$.  Of course,
$\delta(k)$ should be a smooth function and vanish as
$k\to\infty$.  An efficient way to avoid spurious jumps by $\pi$ in
the numerical calculation of $\delta(k)$ is to define
\begin{equation}
    \delta(k,r)=\frac{1}{2i}\, {\rm tr}\, \ln \,
    \left[F^{-1}(r)F^*(r)\right]\ .
    \label{raddelta}
\end{equation}
By construction $\delta(k)=\lim_{r\to0}\delta(k,r)$.  We then consider
\begin{equation}
    \frac{d\delta(k,r)}{dr}
    =\frac{1}{2i}\, {\rm tr}\,
    \left[\frac{d}{dr}F^{*}(F^*)^{-1}
    -\frac{d}{dr}FF^{-1}\right]\, \label{ddelta}
\end{equation}
as an independent function to be included in the numerical routine
that integrates the differential equations for $F$, with the
boundary condition $\lim_{r\to\infty}\delta(k,r)=0$.  Then
$\lim_{r\to0}\delta(k,r)$ will then be a smooth function of $k$ and go
to zero as $k\to\infty$.

We construct the Born series for $F(r)$ in Appendix A. We introduce
$F^{(n_s,n_p)}(r)$, where $n_s$ and $n_p$ label the order in the
expansion around $s(r)=1$ and $p(r)=0$ respectively.  Then
we find for the first two orders
\begin{eqnarray}
    \delta^{(1)}(k)&=&\frac{1}{2i}\lim_{r\to0}\,{\rm tr}
    \left[F^{(1)*}(r)-F^{(1)}(r)\right]
    \label{Bfirst} \\
    \delta^{(2)}(k)&=&\frac{1}{2i}\lim_{r\to0}\,{\rm tr}
    \Big[F^{(2)*}(r)-F^{(2)}(r)-\frac{1}{2}[F^{(1)}(r)]^2
    +\frac{1}{2}[F^{(1)*}(r)]^2\Big]
    \label{Bsec}
\end{eqnarray}
where
\begin{equation}
    F^{(1)}(r)=F^{(1,0)}(r)+F^{(0,1)}(r)\quad {\rm and}\quad
    F^{(2)}(r)=F^{(2,0)}(r)+F^{(0,2)}(r)+F^{(1,1)}(r)\ .
\nonumber \end{equation}
Subtracting these two terms from the full phase shift eliminates
the quadratic divergence from the vacuum polarization energy.
Eliminating the logarithmic divergence would be considerably more
complicated because an expansion up to fourth order in $n_s+n_p$
would be required.\footnote{When restricting to field configurations
with $\Phi\Phi^\dagger=v^2$, two subtractions are
sufficient~\cite{Baa}.}  In Appendix B we introduce a simplified
treatment for the logarithmic divergence. The resulting expression for
the vacuum polarization energy, eq.~(\ref{ecasgeneral}), then becomes
\begin{eqnarray}
    E_{\rm vac} &=&-\frac{1}{2}\sum_j(2G_j+1)|\epsilon_j|
    -\int_0^\infty\frac{dk}{2\pi}\, \sqrt{k^2+m^2}\,
    \frac{d}{dk}\, \bar{\delta}(k)\, \,         
    +E^{(2)}_{\rm ct}+E^{(4)}_{\rm l,ct} 
    \label{evac} \\
    \bar{\delta}(k)&=&\sum_{G,\sigma,\Pi} (2G+1)\left(
    \delta_{G,\sigma,\Pi}(k)-\delta^{(1)}_{G,\sigma,\Pi}(k)
    -\delta^{(2)}_{G,\sigma,\Pi}(k)\right)     \nonumber \\* &&
    {}-m^4\left(\frac{1}{m}\arctan \frac{m}{k}+\frac{k}{k^2+m^2}\right)
    \int_0^\infty dr\,   r^2\left[\left(s(r)^2+p(r)^2-1\right)^2-4(s(r)-1)^2\right]
    \hspace{0.8cm}
    \label{dbar} \\
    E^{(2)}_{\rm ct}&=&\frac{m^2}{\pi^2}\int_0^\infty dq\,   q^2
    \left[h^2(q)+p^2(q)\right]\Big\{q^2+m_H^2
    \nonumber \\ && \hspace{3.0cm}
    {}-6\int_0^1dx\left[m^2+x(1-x)q^2\right]
    \ln \,\frac{m^2+x(1-x)q^2}{m^2-x(1-x)m_H^2}\Big\}
    \nonumber \\ &&    {} -\frac{m^2}{\pi^2}\int_0^\infty dq\,   q^2 p^2(q)
    \Big\{m_H^2+2m^2\int_0^1dx\Big[3\ln \,
    \left(1-x(1-x)\frac{m_H^2}{m^2}\right)
    \nonumber \\ && \hspace{4.0cm}
    {}-2\ln \,\left(1+x(1-x)\frac{q^2}{m^2}\right)\Big]\Big\}
    \label{ect2}\\
    E^{(4)}_{\rm l,ct}&=&\frac{m^4}{8\pi}
    \left(\frac{m_H^2}{m^2}+6\int_0^1dx\,\ln \,
    \left[1-x(1-x)\frac{m_H^2}{m^2}\right]\right)
    \nonumber \\* && \hspace{3.0cm} \times
    \int_0^\infty dr\,   r^2\left[\left(s(r)^2+p(r)^2-1\right)^2-4(s(r)-1)^2\right]
    \label{ect4}
\end{eqnarray}
where the last term in $\bar\delta(k)$ implements the subtraction of the
logarithmic divergence and is compensated by the terms in
$E^{(4)}_{{\rm l,ct}}$ (see Appendix B).  $\sigma$ denotes the sign of
the energy eigenvalue, so that $\omega=\sigma\sqrt{k^2+m^2}$.  We have
introduced the Fourier transforms
\begin{equation}
    h(q)=\int_0^\infty dr\,   r^2\, \frac{\sin qr}{qr}(s(r)-1)
    \quad {\rm and}\quad
    p(q)=\int_0^\infty dr\,   r^2\left[\frac{\cos qr}{qr}
    -\frac{\sin qr}{(qr)^2}\right]p(r)\ .
    \label{ftrans}
\end{equation}

Next we need to find the fermion number carried by the background field. 
We could consider an adiabatic transformation from the trivial background
to the configuration at hand, and count the levels that cross zero energy. 
In general, however, such a procedure is cumbersome.  Instead, we apply the
procedure developed in Ref. \cite{us2}, which is based on Levinson's
theorem.  In each channel, we compare the number of positive energy bound
states $n^{(+)}_{G,\Pi}$ with the number of bound states that have left the
positive continuum $\frac{1}{\pi}\delta_{G,+,\Pi}(0)$.  If one level
originating in the positive energy continuum crosses zero, we will find
that $n^{(+)}_{G,\Pi} = \frac{1}{\pi}\delta_{G,+,\Pi}(0) - 1$.  In this
case, the polarized vacuum carries charge $(2G+1)$ for each species.  In
general, the vacuum charge is given by
\begin{eqnarray}
    Q_{\rm vac}&=&\sum_{G,\Pi}(2G+1)\left[
    \frac{1}{\pi}\delta_{G,+,\Pi}(0)-n^{(+)}_{G,\Pi}\right]
    \nonumber \\
    &=&-\sum_{G,\Pi}(2G+1)\left[
    \frac{1}{\pi}\delta_{G,-,\Pi}(0)-n^{(-)}_{G,\Pi}\right]\ .
    \label{qvac}
\end{eqnarray}
The second equation
reflects the equivalent counting procedure for negative energy
states.  We are interested in configurations with  fermion number~$1$.  If
$Q_{\rm vac}=0$ the fermion number is obtained by explicitly occupying a
level.  
The lowest energy cost arises from filling the
    level with the largest binding.
If  $Q_{\rm vac}=1$ the polarized vacuum already provides the  
fermion number and none of the bound states needs to be explicitly occupied.

\bigskip
\bigskip
\stepcounter{section}
\leftline{\large\bf 4. Numerical Analysis}
\bigskip

Our formalism is set up to allow consideration of an arbitrary
background $\Phi(\vec x)$ of the form~(\ref{background}). However,
as in all variational methods, we limit ourselves to variation of a few
parameters in an {\it ansatz} motivated by physical considerations.
We will scale energies and lengths in terms of the fermion mass $m$.
We choose a four parameter {\it ansatz}
\begin{eqnarray}
    s+i\vec{\tau}\cdot\vec{p}&=&\rho(\xi)\, {\rm
    exp}\,\left(i\vec{\tau}\cdot\hat{x}\Theta(\xi)\right) \nonumber \\
    \rho(\xi)&=&1+b_1\left[1+b_2^2\frac{\xi}{w}\right]
    {\rm exp}\left(-b_2^2\frac{\xi}{w}\right)
    \nonumber \\
    \Theta(\xi)&=&-\pi\frac{{\rm e}^{c^2}-1}
    {{\rm e}^{c^2}-3+2{\rm e}^{c^2\xi/w}}
    \label{ansatz1}
\end{eqnarray}
where $\xi=mr$, and the variational parameters are $w$, $b_1$, $b_2$ and
$c$.  Note that $\Theta(0)=-\pi$, $\rho(0)=1+b_1$, and both $\rho-1$ and
$\Theta$ go to zero exponentially as $\xi\to\infty$, since we expect a
Yukawa tail.  As long as $\rho$ does not vanish, this background has
winding number one. The width $w$ is chosen such that $\Theta(w)=-\pi/3$. 
Furthermore, we have ensured that $\frac{d}{d\xi}\rho(\xi)\big|_{\xi=0}=0$,
as the classical equations of motion require.

In terms of $\rho(\xi)$ and $\theta(\xi)$, the classical energy
eq.~(\ref{ecl1}) is
\begin{eqnarray}
     \frac{1}{m}E_{\rm
    cl}(w,b_1,b_2,c)&=& \frac{2\pi}{g^2}\int_0^\infty d\xi \xi^2 \left\{
    \left(\frac{d\rho}{d\xi}\right)^2+
    \left(\rho\frac{d\Theta}{d\xi}\right)^2 +\frac{2}{\xi^2}{\rm
    sin}^2\Theta +\frac{\mu_H^2}{4}\left(\rho^2-1\right)^2\right\}
    \nonumber \\
    &\equiv& \frac{1}{g^2}{\cal E}_{\rm cl}(w,b_1,b_2,c)
    \label{eclscal}
\end{eqnarray}
where $\mu_H=\frac{m_H}{m}$.  Then the total energy of the
configuration with fermion number $1$ is
\begin{equation}
    \frac{1}{m}E_{\rm tot}=
    \frac{1}{g^2}{\cal E}_{\rm cl}(w,b_1,b_2,c)
    +\left(1-Q_{\rm vac}\right)
    \epsilon_1(w,b_1,b_2,c)+{\cal E}_{\rm vac}(w,b_1,b_2,c)
    \label{etot}
\end{equation}
where ${\cal E}_{\rm vac}=E_{\rm vac}/m$, and $\epsilon_1=\omega_1/m$
is the energy eigenvalue of the most strongly bound state.
Note that for fixed $\mu_H$, the coupling $g$ appears only in the
coefficient of the classical term.

Configurations with $\Theta(0)=-\pi$ and $\Theta(\infty)=0$ tend to
strongly bind a state originating from the positive continuum in the
$G^{\Pi}=0^{+}$ channel. For $w$ large enough, this bound state will
have crossed zero, causing the polarized vacuum charge to be $Q_{\rm
vac}=1$.  In that case the level is not explicitly occupied and the
corresponding term drops out of eq.~(\ref{etot}).

\subsection{Sample Numerical Calculations}

\begin{figure}[t]
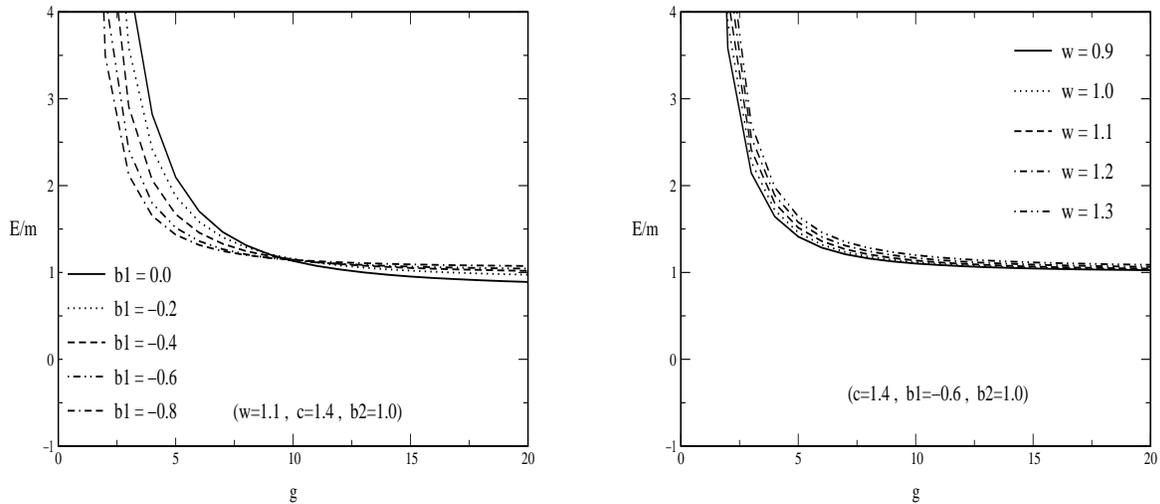

\centerline{
\epsfig{figure=p1noN.eps,height=7.0cm,width=7.0cm,angle=270}
\hspace{1cm}
\epsfig{figure=p2noN.eps,height=7.0cm,width=7.0cm,angle=270}}
\caption{\label{fig_1}\sl The total energy as a function
of the Yukawa coupling constant $g$ with $m_H=0.35v$.}
\end{figure}

For a given set of variational parameters, we first compute the
phase shifts and perform the subtractions according to
eq.~(\ref{dbar}), which allows us to carry out the momentum integral in
eq.~(\ref{evac}).  Using Levinson's theorem, we then find the number of bound
states in a given channel from $\delta_{G,\sigma,\Pi}(k=0)$, and use
shooting to compute the bound state energies $\epsilon_j$ once we
know how many to look for.  In terms of the scaled
variables, the bound state and phase shift contributions in
eq.~(\ref{evac}) depend on {\it ansatz} parameters, but not on model
parameters.  The dependence on model parameters is completely
contained in ${\cal E}_{\rm cl}/g^2$, ${\cal E}_{\rm ct}^{(2)}=E_{\rm
ct}^{(2)}/m$ and ${\cal E}_{\rm l,ct}^{(4)}=E_{\rm l,ct}^{(4)}/m$,
which are simple integrals involving the background fields.  They can
easily be obtained numerically for large regions in the
model parameter space.  Hence an efficient strategy is to choose
a set of variational parameters and then consider the total
energy as a function of the model parameters for that particular
background configuration.  In Fig.~\ref{fig_1} we display a typical
result for the total energy as a function of the Yukawa coupling $g$.

The existence of configurations with total energy $E_{\rm tot}/m<1$
shows that there is a stable soliton whose energy is at most $E_{\rm
tot}$.  Apparently a sizable Yukawa coupling $g$ is needed to
obtain a stable soliton. However, as we will discuss later, our model
is not reliable for such large Yukawa couplings because the Landau
pole appears at an energy scale comparable to $1/w$.

\begin{figure}[hbt]
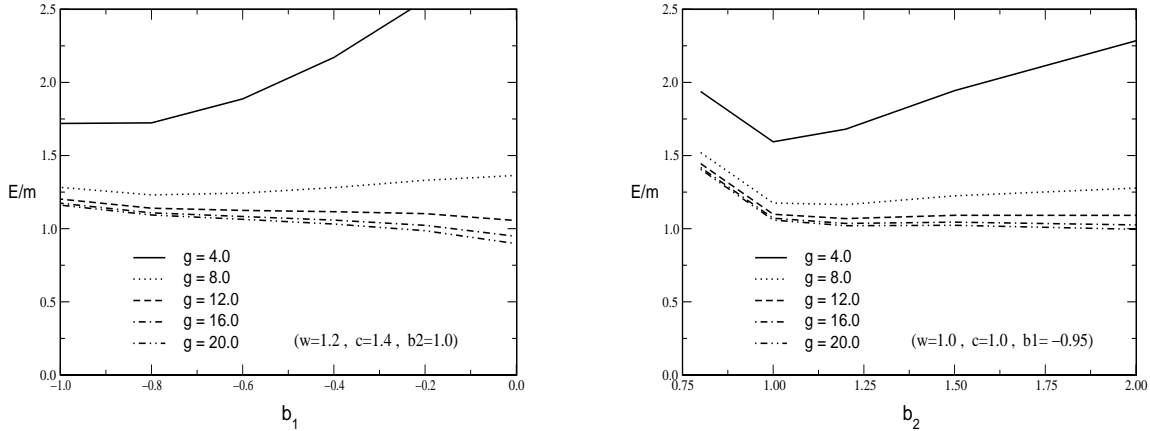

\centerline{
\epsfig{figure=p3anoN.eps,height=7.0cm,width=6.0cm,angle=270}
\hspace{1cm}
\epsfig{figure=p3bnoN.eps,height=7.0cm,width=6.0cm,angle=270}}
\caption{\label{fig_2}\sl Total energy as a function of the depth
parameters $b_1$ and $b_2$ with $m_H = 0.35 v$ and for various values of
the Yukawa coupling constant $g$.}
\end{figure}

In Fig.~\ref{fig_2} we display the total energy as a function
of the depth parameters $b_1$ and $b_2$, for various values of the
Yukawa coupling constant $g$ and typical values of the remaining
variational parameters $w$ and $c$.
We observe a shallow local minimum in the vicinity of $b_1=-0.8$ for
small and moderate values of $g$.  However, at this minimum the total
energy is larger than the mass $m$ of the free fermion.  For larger
$g$, we obtain a total energy less than $m$ for configurations with
$b_1\approx 0$.  Configurations with $b_1>0$ are more strongly bound,
but the one fermion loop approximation fails for such configurations
because the energy functional is not bounded from below (see
Appendix~C).  We have therefore restricted the space of variational
parameters to configurations for which the vacuum is stable to one loop.

We observe a similar behavior for $E_{\rm tot}$ as a function of
$b_2$. There exists a local minimum for small and moderate $g$ that does not
yield a bound soliton.  For large $g$, a bound soliton seems possible
if $b_2$ is big enough.  In this case, the vacuum is stable
at one loop for these values of the variational parameters.  Note that
when we find a marginally bound configuration, the vacuum polarization
contribution to the energy $E_{\rm vac}$ tends to almost exactly
compensate for the gain from binding a single level.

\subsection{Comparison with the Derivative Expansion}

In order to check our computation of the vacuum polarization energy, in
particular the simplified treatment of the logarithmic divergence, we have
compared our results with the derivative expansion.  The relevant formulas
for the derivative expansion are provided in Appendix C. Denoting by ${\cal
E}_{\rm grad}$ the energy computed to second order in the derivative
expansion as obtained from eq.~(\ref{gradres}), we display
\begin{equation}
     \Delta_{1} \equiv\frac{{\cal E}_{\rm cl}+
     {\cal E}_{\rm vac}-{\cal E}_{\rm grad}}
     {{\cal E}_{\rm cl}+{\cal E}_{\rm vac}+{\cal E}_{\rm grad}}
     \label{grad1}
\end{equation}
as a function of the width parameter $w$.  The other variational
parameters are kept constant at $c=1.0$, $b_1=-0.4$ and $b_2=1.0$.
Also, we consider various values for the coupling constant $g$.
\begin{table}
\caption{\label{tab_1}Comparison of the classical and
renormalized vacuum polarization energy with the derivative 
expansion, {\it cf.}\ eq.~(\protect\ref{grad1}).}
\vskip0.2cm
\centerline{
\begin{tabular}{ccccc}
$w$ & 1.0 & 2.0 & 2.5 & 3.0 \\
\hline
$g= \phantom{0}5.0$ & --0.127 & --0.024 & --0.008 & 0.003\\
$g=10.0$ & --0.240 & --0.041 & --0.013 & 0.005\\
$g=15.0$ & --0.287 & --0.048 & --0.015 & 0.007\\
$g=20.0$ & --0.308 & --0.050 & --0.015 & 0.006 
\end{tabular}}
\end{table}

Where the derivative expansion is valid we expect $\Delta_1$ to go to
zero.  From Table~\ref{tab_1}, we conclude that our calculation agrees
with the derivative expansion at large $w$, where the derivative
expansion becomes exact.  In this region, the derivative expansion is a
good check on our computation of the vacuum polarization energy.  In
particular, it checks our handling of renormalization, since 
renormalization effects are included in the derivative expansion in the
standard way.  On the other hand, it is clear from Table~\ref{tab_1}
that the second-order derivative expansion cannot be trusted for
$w\approx1.0$, {\it i.e.\/} for background configurations whose
extension is close to the Compton wavelength of the fermion.
One might imagine improving the derivative expansion by including the
effect of an explicitly occupied level,  
\begin{equation}
    \Delta_{2}\equiv\frac{{\cal E}_{\rm tot}-{\cal E}_{\rm grad}}
    {{\cal E}_{\rm tot}+{\cal E}_{\rm grad}}
    \label{grad2}
\end{equation}
because the derivative expansion to $Q_{\rm vac}$, the
topological charge, suggests that the background configuration carries
fermion number regardless of the value for $w$.  However, this modification
does not give any better agreement at  small $w$, as can be observed from
Table~\ref{tab_2}.  In general, the inclusion of the explicitly occupied
level tends to change the sign of the relative deviation.
\begin{table}
\caption{\label{tab_2}Comparison of the total
energy with the gradient expansion, {\it cf.}\ eq.~(\protect\ref{grad2}).}
\vskip0.2cm
\centerline{
\begin{tabular}{ccccc}
$w$ & 1.0 & 2.0 & 2.5 & 3.0 \\
\hline
$g= \phantom{0}5.0$ & 0.168 & 0.034 & 0.018 & 0.012\\
$g=10.0$ & 0.257 & 0.056 & 0.029 & 0.019\\
$g=15.0$ & 0.285 & 0.063 & 0.033 & 0.022\\
$g=20.0$ & 0.297 & 0.066 & 0.034 & 0.023 
\end{tabular}}
\end{table}

\subsection{The Landau Pole}

From these studies one might conclude that a soliton takes over the
role of the lightest fermion once the Yukawa coupling constant $g$
becomes large enough.  At large $g$, the positive contribution to the total
energy from ${\cal E}_{\rm cl}$ in eq.~(\ref{eclscal}),
which disfavors the soliton, decreases quickly for large $g$.
However, for large couplings the model itself becomes ill-defined.
Since the model is not asymptotically free, it has a Landau
singularity in the ultraviolet, reflecting new dynamics at some cutoff
scale.  Thus the Landau pole sets a minimum distance scale below which
the model is not consistent.  Solitons that are large compared
to this scale are relatively insensitive to the unknown dynamics at
the cutoff scale, but solitons whose size is comparable to this scale
cannot be trusted.  In this section we will discuss the emergence of
the Landau pole and estimate its effect on the vacuum polarization
energy by comparing the present results with a calculation that
removes this pole.  Although this removal is somewhat {\it ad hoc\/},
it nevertheless provides some insight into the reliability of the
computations in case of large $g$.

Denoting the Fourier transforms of $s(r)-1$ and $\vec{p}(r)$ by $h(\vec{q})$
and $\vec{p}(\vec{q})$, respectively, the contribution of the
two-point function to the total energy can be written as
\begin{equation}
    E_2=\frac{v^2}{2}\int \frac{d^3q}{(2\pi)^2}
    \left\{G_h^{-1}(-\vec{q}\,^2)h(\vec{q})h(-\vec{q})
    +G_p^{-1}(-\vec{q}\,^2)\vec{p}(\vec{q})\cdot\vec{p}(-\vec{q})\right\}
    \label{etwopt}
\end{equation}
where
\begin{eqnarray}
    G_h^{-1}(q^2)&=&\frac{v^2}{2}\Bigg\{
    \left(q^2-m_H^2\right)\left(1+\frac{g^2}{4\pi^2}\right)
    \nonumber \\ &&
    +\frac{g^2}{4\pi^2}6\int_0^1dx
    \left[m^2-x(1-x)q^2\right]\ln 
    \frac{m^2-x(1-x)q^2}{m^2-x(1-x)m_H^2}\Bigg\}
    \label{h2ptfct} \\
    G_p^{-1}(q^2)&=&\frac{v^2}{2}\Bigg\{
    q^2\left(1+\frac{g^2}{4\pi^2}\right)
    +\frac{g^2}{4\pi^2}\Bigg[
    6\int_0^1dx
    \left[m^2-x(1-x)q^2\right]\ln 
    \frac{m^2-x(1-x)q^2}{m^2-x(1-x)m_H^2}
    \nonumber \\ &&
    +2m^2\int_0^1dx\left[3\ln \left(
    1-x(1-x)\frac{m_H^2}{m^2}\right)
    -2\ln \left(
    1-x(1-x)\frac{q^2}{m^2}\right)\right]\Bigg]\Bigg\}
    \label{p2ptfct}
\end{eqnarray}
which includes classical, loop, and counterterm contributions.
$G_p(q^2)$ has a pole (the Landau ghost pole) at space-like
momentum $-m_G^2$ with residue $Z_G$. The pole location is
easily obtained numerically from the condition
$G_p^{-1}(q^2=-m_G^2)=0$.  In the vicinity of $q^2\approx-m_G^2$ we
have the expansion
\begin{equation}
    G_p^{-1}(q^2)=\frac{1}{Z_G}\left(q^2+m_G^2\right)
    +{\cal O}\left(q^2+m_G^2\right)^2
    \label{deflandau}
\end{equation}
with
\begin{eqnarray}
    \frac{1}{Z_G}&=&\frac{\partial}{\partial q^2}G_p^{-1}(q^2)
    \Big|_{q^2=-m_G^2}
    \nonumber \\
    &=&\frac{v^2}{2}\Bigg\{1-\frac{g^2}{4\pi^2}\Bigg[
    6\int_0^1dxx(1-x)\ln 
    \frac{m^2+x(1-x)m_G^2}{m^2-x(1-x)m_H^2}
    \nonumber \\ && \hspace{3.0cm}
    -4\int_0^1dx\frac{x(1-x)}{m^2+x(1-x)m_G^2}\Bigg]\Bigg\}\ .
\label{residuum}
\end{eqnarray}
The existence of this pole yields an unphysical negative contribution
to the total energy at large space-like momenta, or equivalently for
narrow background field configurations.  Based on the
K\"all\'en-Lehmann representation for the two-point function, the
authors of Ref.~\cite{Ha94} suggested a procedure to eliminate the
Landau pole while maintaining chiral symmetry.  They replace
eq.~(\ref{etwopt}) with
\begin{equation}
    \bar{E}_2=\frac{v^2}{2}\int \frac{d^3q}{(2\pi)^2}
    \Delta_p^{-1}(-\vec{q}\,^2)\left\{h(\vec{q})h(-\vec{q})
    +\vec{p}(\vec{q})\cdot\vec{p}(-\vec{q})\right\}
    \label{ewopole}
\end{equation}
where $\Delta_p(q^2)=G_p(q^2)-Z_G/(q^2+m_G^2)$ removes the Landau
pole.  We can easily adopt this procedure since we have already
extracted the loop and counterterm contributions from the two-point
function in eq.~(\ref{ect2}).  That is, we replace $E_{\rm cl}+E_{\rm
ct}^{(2)}+E_{\rm l,ct}^{(4)}$ by $\bar{E}_2+E_{\rm cl}^{(3,4)}$ with
\begin{equation}
    \frac{1}{m}E_{\rm cl}^{(3,4)}
    =\frac{\pi \mu_H^2}{2g^2}\int_0^\infty dx x^2
    \left\{\left[2(s-1)+(s-1)^2+\vec{p}\,^2\right]^2-4(s-1)^2\right\}\ .
    \label{ecl34}
\end{equation}
\begin{figure}[hbt]
\centerline{
\epsfig{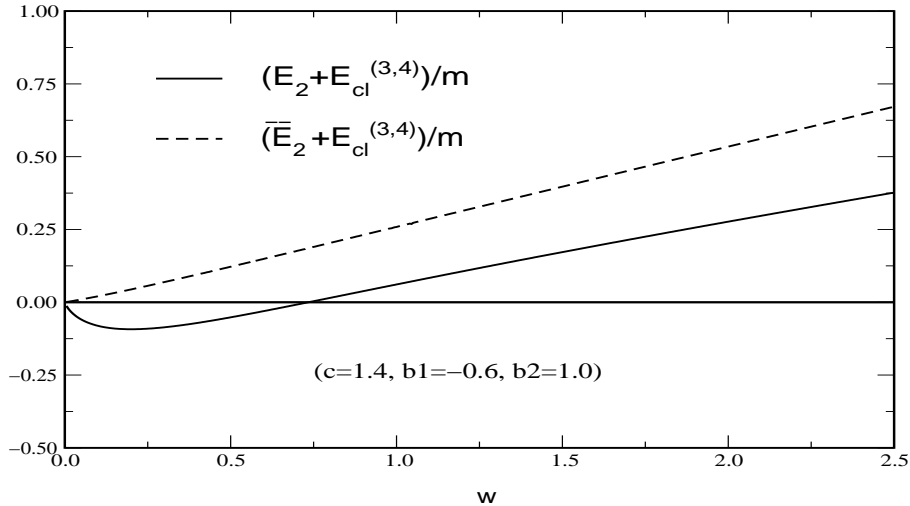}}
\caption{\label{fig_3}\sl The Landau pole subtraction, for $m_H =0.35v$
and $g=12$.}
\end{figure}%
In Fig.~\ref{fig_3} we show the effect of this replacement as a
function of the width parameter $w$ for $g=12$, which
is in the region where a bound soliton can occur.  For small $w$, we
observe that the original computation,
eq.~(\ref{etwopt}), gives a negative contribution.  However, for small
$w$ there are only weakly bound states and the vacuum is almost
undistorted.  The classical energy is also small since $g$ is large.
Hence the total energy is dominated by the renormalized Feynman
diagram contribution $E_{\rm ct}^{(2)}+E_{\rm l,ct}^{(4)}$, which can
be negative due to the Landau pole.  Thus for small $w$ and large $g$,
the Landau pole dominates the binding of the soliton, and, even
worse, the total energy could be negative, reflecting an unphysical
vacuum instability~\cite{Ri87}.  Using the above prescription to
eliminate this pole, the total energy turns out to be positive for all
values of $w$, so the instability is removed.  As can be seen from
Fig.~\ref{fig_3}, for sensible $w$ this prescription increases the
total energy by about $0.25m$ for the parameters chosen, which
unbinds the soliton.  We conclude that the solitons found at large $g$
are principally bound by unphysical effects associated with the Landau
singularity, and not by reliable dynamical properties of the model.

\subsection{Scalar Backgrounds}

\begin{figure}[hbt]
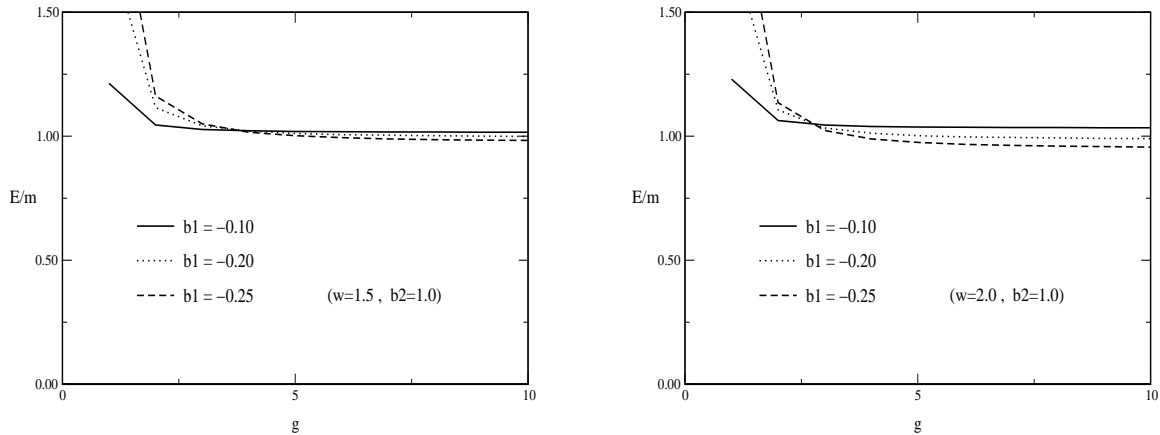

\centerline{
\epsfig{figure=p5noN.eps,height=7.0cm,width=6.0cm,angle=270}
\hspace{1cm}
\epsfig{figure=p6noN.eps,height=7.0cm,width=6.0cm,angle=270}}
\caption{\label{fig_4}\sl The total energy with only the scalar background 
as a function of the Yukawa coupling constant $g$ with $m_H=0.35v$. Note
that the variational parameter~$c$ is irrelevant without pseudoscalar fields.}
\end{figure}

We conclude the numerical analysis with a calculation of the total energy
when only a scalar background field $s(r)$ is present.  Our goal is a
brief comparison with the results of Ref.~\cite{Ba91}, rather than a
complete study.  For $p\equiv0$ the Dirac Hamiltonian is charge
conjugation invariant.  Hence the charge carried by the background
field is zero and we must explicitly occupy the most strongly bound
state.  In Fig.~\ref{fig_4} we show typical results of the numerical
calculation for the total energy as a function of the coupling constant.
The figure shows that a slightly bound soliton emerges
even for modest values of the Yukawa coupling.  Its energy is
up to 5\% less than that of a fermion in the background of the
translationally invariant scalar field.  In this case, the Landau
ghost singularity in $G_p$ ({\it cf.}\ eq.~(\ref{p2ptfct})) is
irrelevant and we may trust the calculation even for small
$w$. Furthermore, the second-order derivative expansion deviates from
the full calculation by only a fraction of a percent even at moderate
values of the width parameter, $w\approx1.5$.  Plotting the
results from the derivative expansion would yield indistinguishable
curves in Fig.~\ref{fig_4}.  Our approach thus confirms the findings of
Ref. \cite{Ba91}, which used the derivative expansion to find a slightly
bound soliton.

\bigskip
\bigskip
\stepcounter{section}
\leftline{\large\bf 5. Summary and Outlook}
\bigskip

We have performed a quantitative search for Higgs solitons in a theory with
chiral fermions. In the analogous model in one spatial dimension, such
solitons exist and are strongly bound for a wide range of coupling
constants.  In three spatial dimensions, however, we did not find any
region where the soliton binding was strong enough to be convincing. 
Twisted Higgs background fields do strongly bind a fermion level, but it is
necessary to add the renormalized energy due to the distortion of the
fermion vacuum.  We have developed new methods to regularize, renormalize,
and compute the corresponding vacuum polarization energy.  In this model,
we find that the vacuum polarization tends to cancel the contribution of
the strongly bound fermion level.  The total one-loop energy overcomes the
classical energy only for large  Yukawa couplings, where the theory has an
unphysical Landau pole.  Hence energetically favored  Higgs solitons
observed for large Yukawa couplings cannot be interpreted as reliable
predictions of the model.

Nevertheless, these findings do not rule out
the existence of Higgs solitons within the Standard Model.  In
particular, the inclusion of the gauge fields could be critical to
soliton formation.  This point of view is motivated by a number of
considerations:

\begin{itemize}
\item
Expanding the variational {\it ansatz} to include gauge fields can
only decrease the energy of a configuration.  In this respect, the
$3+1$ dimensional model is different from the $1+1$ dimensional model
where we did find a soliton, because in one dimension gauge fields
do not add any new interactions. 

\item
Gauge fields are essential to the anomaly arguments underlying the
decoupling results.

\item
In the gauge theory, there is a sphaleron barrier with height $2\pi
v/g_W$, where $v$ is the Higgs vacuum expectation value and $g_W$ is
the gauge coupling.  If the fermion has a Yukawa coupling $g_F$ such
that its perturbative mass $g_F v$ is much larger than the sphaleron
height, it has an unsuppressed decay mode over the sphaleron.  Thus
the ordinary fermion states are unstable and are nowhere to be found
in the spectrum of the fermion Hamiltonian.  The creation of a soliton
state with mass below the sphaleron energy would allow the theory to
remain gauge invariant after decoupling as required by \cite{decoupling}.

\end{itemize}

Work is now underway to generalize this calculation to include the
gauge fields.

\bigskip
\bigskip
\leftline{\large\bf Acknowledgments}
\bigskip

We thank Vishesh Khemani for helpful discussions.  E.F. and
R.L.J. are supported in part by the U.S.~Department of Energy
(D.O.E.) under cooperative research agreement~\#DF-FC02-94ER40818.
N.G. is supported by the U.S.~Department of Energy (D.O.E.) under
cooperative research agreement~\#DE-FG03-91ER40662 and H.W. is
supported by the Deutsche Forschungsgemeinschaft under contracts We
1254/3-1,2.

\appendix

\bigskip
\bigskip
\stepcounter{chapter}
\leftline{\large\bf Appendix A: Dirac equations}
\bigskip

In this Appendix we present the first- and second-order Dirac equations
used in Section~3 for fermions in a static background field in the
spherical {\it ansatz}, eq.~(\ref{background}).  In order to solve the
Dirac equation, eqs.~(\ref{dirham},\ref{direq}), we begin with spinors
that are eigenstates of parity and total grand spin \cite{Al96}, where
grand spin $\vec G$ is the sum of total angular momentum
$\vec{j}=\vec{l}+\frac{1}{2}\vec{\sigma}$ and isospin $\half
\vec\tau$.  For a given grand spin quantum number $G$ with
$z$-component $M$, we will find the bound state and scattering
wavefunctions in terms of the spherical harmonic functions ${\cal
Y}_{j,\ell}(\hat{x})$ with $j=G\pm\frac{1}{2}$ and
$\ell=j\pm\frac{1}{2}$.  These are two-component spinors in both spin
and isospin space.  In the following we will suppress the label $M$
for the $z$-component.  While grand spin is conserved, the background
field in eq.~(\ref{background}) mixes states with different total angular
momentum $j$. For the channels with parity $\Pi=+(-)^G$, the spinor
that diagonalizes eq.~(\ref{dirham}) reads
\begin{equation}
    \Psi^{(+)}_G(\vec{x})=
    \pmatrix{ig_1(r){\cal Y}_{G+\frac{1}{2},G}(\hat x)\cr
    f_1(r){\cal Y}_{G+\frac{1}{2},G+1}(\hat x) } +
    \pmatrix{ig_2(r){\cal Y}_{G-\frac{1}{2},G}(\hat x)\cr
    -f_2(r){\cal Y}_{G-\frac{1}{2},G-1}(\hat x)}\ .
    \label{pos1}
\end{equation}
The spinor with opposite parity, $-(-)^G$, is parameterized as
\begin{equation}
    \Psi^{(-)}_G(\vec{x})=
    \pmatrix{ig_1(r){\cal Y}_{G+\frac{1}{2},G+1}(\hat x)\cr
    -f_1(r){\cal Y}_{G+\frac{1}{2},G}(\hat x)}+
    \pmatrix{ig_2(r){\cal Y}_{G-\frac{1}{2},G-1}(\hat x)\cr
    f_2(r){\cal Y}_{G-\frac{1}{2},G}(\hat x)}\ .
    \label{pos2}
\end{equation}
Note that for convenience we have again suppressed both the grand spin
and parity labels for the radial functions $g_i$ and $f_i$.

The matrix elements of the operator $\vec{\tau}\cdot\hat{x}$ between
various ${\cal Y}_{jl}$ can be found in the
literature~\cite{Ka84}.  Then the Dirac equation can be expressed as
coupled first-order differential equations for the radial functions
$g_i$ and $f_i$.  In the $\Pi=+(-)^G$ parity channel, we have
\begin{eqnarray}
    0&=&g_1^\prime-\frac{G}{r}g_1+(ms+\omega)f_1
    -\frac{mp}{2G+1}\left(g_1-2\sqrt{G(G+1)}g_2\right)
    \nonumber \\
    0&=&g_2^\prime+\frac{G+1}{r}g_2-(ms+\omega)f_2
    +\frac{mp}{2G+1}\left(g_2+2\sqrt{G(G+1)}g_1\right)
    \nonumber \\
    0&=&f_1^\prime+\frac{G+2}{r}f_1+(ms-\omega)g_1
    +\frac{mp}{2G+1}\left(f_1+2\sqrt{G(G+1)}f_2\right)
    \nonumber \\
    0&=&f_2^\prime-\frac{G-1}{r}f_2-(ms-\omega)g_2
    -\frac{mp}{2G+1}\left(f_2-2\sqrt{G(G+1)}f_1\right)
    \label{1stpos}
\end{eqnarray}
where a prime indicates a derivative with respect to $r$.  In the
$\Pi=-(-)^G$ parity channel, we have
\begin{eqnarray}
    0&=&g_1^\prime+\frac{G+2}{r}g_1-(ms+\omega)f_1
    -\frac{mp}{2G+1}\left(g_1-2\sqrt{G(G+1)}g_2\right)
    \nonumber \\
    0&=&g_2^\prime-\frac{G-1}{r}g_2+(ms+\omega)f_2
    +\frac{mp}{2G+1}\left(g_2+2\sqrt{G(G+1)}g_1\right)
    \nonumber \\
    0&=&f_1^\prime-\frac{G}{r}f_1-(ms-\omega)g_1
    +\frac{mp}{2G+1}\left(f_1+2\sqrt{G(G+1)}f_2\right)
    \nonumber \\
    0&=&f_2^\prime+\frac{G+1}{r}f_2+(ms-\omega)g_2
    -\frac{mp}{2G+1}\left(f_2-2\sqrt{G(G+1)}f_1\right)\ .
    \label{1stneg}
\end{eqnarray}
We can use these differential equations to obtain the bound state
solutions with $|\omega|< m$ using ordinary shooting methods.  For a
number of cases, we have verified the solutions numerically by
diagonalizing eq.~(\ref{dirham}) in a spherical cavity \cite{Al94}. As
discussed in Section 3, we also require the second-order equations
for either the upper ($g_i$) or lower ($f_i$) components of the Dirac
spinor.  These components will be collected into the matrix
$F(r)H(kr)$ defined in eq.~(\ref{defF}).  In Section~3 we chose to
work with the upper components only. Here we will discuss both upper
and lower components, $F_{u}(k)$ and $F_{l}(k)$. Substituting this
parameterization into the second-order equations for the
upper or lower components of the Dirac equation and multiplying by the
inverse of $H(kr)$ from the right, we obtain for the upper components
in the channel with $\Pi=+(-)^G$ %
\begin{equation}
    0=F_u^{\prime\prime}+\frac{2}{r}F_u^\prime
    \left[1+rL_+^\prime(kr)\right]-\frac{s^\prime}{s+\frac{\omega}{m}}
    \left[F_u^\prime+F_u L_+^\prime(kr)\right]+V(r,m) F_u 
    \label{fup}
\end{equation}
with $F=F_u$ entering eq.~(\ref{Fdifffeq}).  The matrix $L_+(kr)$ is
defined in eq.~(\ref{defL}).  For later convenience we have added the
fermion mass $m$ as an argument of the potential matrix,
\begin{eqnarray}
    V_{11}(r,m)&=&-m^2\left[s^2+p^2-1\right]
    +\frac{G}{r}\frac{s^\prime}{s+\frac{\omega}{m}}
    -\frac{mp^\prime}{2G+1}
    -\frac{2}{r}\frac{(G+1)mp}{2G+1}
    +\frac{mp}{2G+1}\frac{s^\prime}{s+\frac{\omega}{m}}
    \nonumber \\
    V_{22}(r,m)&=&-m^2\left[s^2+p^2-1\right]
    -\frac{G+1}{r}\frac{s^\prime}{s+\frac{\omega}{m}}
    +\frac{mp^\prime}{2G+1}
    -\frac{2}{r}\frac{Gmp}{2G+1}
    -\frac{mp}{2G+1}\frac{s^\prime}{s+\frac{\omega}{m}}
    \nonumber \\
    V_{12}(r,m)&=&V_{21}(r,m)=\frac{2\sqrt{G(G+1)}}{2G+1}
    \left[mp^\prime+mp\left(\frac{1}{r}-
    \frac{s^\prime}{s+\frac{\omega}{m}}\right)\right]\ .
    \label{defV}
\end{eqnarray}
Similarly, we find the second-order equation for the
lower components in the channel with $\Pi=+(-)^{G}$ is
\begin{equation}
    0=F_l^{\prime\prime}+\frac{2}{r}F_l^\prime
    \left[1+rL_-^\prime(kr)\right]
    -\frac{s^\prime}{s-\frac{\omega}{m}}
    \left[F_l^\prime+F_l L_-^\prime(kr)\right]
    +W(r,m) F_l -\left[K,F_l\right]
    \label{flp}
\end{equation}
where
\begin{eqnarray}
    W_{11}(r,m)&=&-m^2\left[s^2+p^2-1\right]
    -\frac{G+2}{r}\frac{s^\prime}{s-\frac{\omega}{m}}
    +\frac{mp^\prime}{2G+1}
    -\frac{2}{r}\frac{(G+1)mp}{2G+1}
    -\frac{mp}{2G+1}\frac{s^\prime}{s-\frac{\omega}{m}}
    \nonumber \\
    W_{22}(r,m)&=&-m^2\left[s^2+p^2-1\right]
    +\frac{G-1}{r}\frac{s^\prime}{s-\frac{\omega}{m}}
    -\frac{mp^\prime}{2G+1}
    -\frac{2}{r}\frac{Gmp}{2G+1}
    +\frac{mp}{2G+1}\frac{s^\prime}{s-\frac{\omega}{m}}
    \nonumber \\
    W_{12}(r,m)&=&W_{21}(r,m)=\frac{2\sqrt{G(G+1)}}{2G+1}
    \left[mp^\prime-mp\left(\frac{1}{r}+
    \frac{s^\prime}{s+\frac{\omega}{m}}\right)\right]\ .
    \label{defW}
\end{eqnarray}
In this case we have two different orbital angular momenta, leading to
the commutator term $K=(1/r^2)\,{\rm diag}\,\{(G+1)(G+2),G(G-1)\}$.
Next we write down the differential equations in the channel with
$\Pi=-(-)^G$ .  Using definitions analogous to eqs.~(\ref{defF})
we find for the upper components
\begin{equation}
    0=F_u^{\prime\prime}+\frac{2}{r}F_u^\prime
    \left[1+rL_-^\prime(kr)\right]-\frac{s^\prime}{s+\frac{\omega}{m}}
    \left[F_u^\prime+F_u L_-^\prime(kr)\right]
    +W(r,-m) F_u -\left[K,F_u\right]
    \label{fun}
\end{equation}
while the lower components obey %
\begin{equation}
    0=F_l^{\prime\prime}+\frac{2}{r}F_l^\prime
    \left[1+rL_+^\prime(kr)\right]-\frac{s^\prime}{s-\frac{\omega}{m}}
    \left[F_l^\prime+F_lL_+^\prime(kr)\right] +V(r,-m) F_l\ .
    \label{fln}
\end{equation}

We must study the Born series in order to obtain the subtraction terms
in eq.~(\ref{dbar}).  We expand around the free solution,
\begin{equation}
    F_u(r) = 1 + g_s F_u^{(1,0)} (r) + g_p F_u^{(0,1)} (r)
    +g_s^2 F_u^{(2,0)} (r) + g_p^2 F_u^{(0,2)} (r) +
    g_sg_pF_u^{(1,1)} (r)\ldots
    \label{BornA1}
\end{equation}
and similarly for $F_l$. The expansion parameters $g_{p}$ and $g_{s}$
are defined by
\begin{equation}
    \Phi(\vec{x})=v\left[1+g_s(s(r)-1)+
    ig_p\vec{\tau}\cdot\hat{x}\, p(r)\right]
    \label{BornA2}
\end{equation}
where we have expanded the equations of motion and the potential matrices
$V$ and $W$ in terms of the artificial coupling constants $g_s$ and
$g_p$.  Having obtained the second-order Dirac equations at the
desired order in the couplings, we also expand the defining equation
for the phase shifts, eq.~(\ref{delta}), in these constants.  Then,
for example, $F^{(1,0)}$ and $F^{(0,1)}$ will contribute to
$\delta^{(1)}_{G,p,\pi}$.  However, we observe the relations
\begin{equation}
    V^{n,2i+1}(r,m)\longleftrightarrow -V^{n,2i+1}(r,-m) \quad {\rm
    and} \quad W^{n,2i+1}(r,m)\longleftrightarrow -W^{n,2i+1}(r,-m)
    \label{VWsym}
\end{equation}
when $\omega \longleftrightarrow -\omega$.  As a result, once we sum
over parity channels, the positive and negative energy pieces
in the vacuum polarization energy calculation cancel for all odd
powers of the pseudoscalar field.  Of course, this cancellation just
reflects parity invariance.

For scattering solutions, we have $|\omega|>m$ and hence the
second-order equations are nonsingular as long as
$|s|<1$.  As explained in the main part of the paper we integrate
these second-order equations from $r\to\infty$ to $r=0$.  At
$r\to\infty$ we have $s=1$ and commonly $s$ changes sign at some
intermediate point, say $r_0$.  For $\omega>0 $ it is hence
appropriate to use the second-order differential equations for the
upper components, which will be the larger ones.  Eventually, however,
$s$ may become less than minus one.  In that case we will have
singularities when using the upper components. Then it would be more
appropriate to employ the lower components as these will be the larger
ones for $r<r_0$.  For $\omega<0$ the situation is reversed.  We
therefore switch between these two components at $r=r_0$ using the
first-order Dirac equations.  In the parity $(-)^G$ channel the
switch from upper to lower components is given by %
\begin{eqnarray}
    F_l&=&\frac{-1}{\omega+ms}\sigma_3\left(F^\prime_u
    +F_uL_+^\prime+MF_u\right) H_+H_-^{-1}
    \label{sw1}\\
    F_l^\prime&=&-F_lL_-^\prime-\bar{M}F_l
    +(\omega-ms)\sigma_3F_uH_+H_-^{-1}\ .
    \label{sw2}
\end{eqnarray}
In this case the upper components are obtained by integrating from
infinity to $r_0$ and $F_l$ in eq.~(\ref{sw2}) is given by
eq.~(\ref{sw1}). The matrices $M$ and $\bar M$ are defined by
\begin{equation}
    M=\pmatrix{-\frac{G}{r}-\frac{mp}{2G+1} &
    2\frac{\sqrt{G(G+1)}}{2G+1}mp \cr
    2\frac{\sqrt{G(G+1)}}{2G+1}mp & \frac{G+1}{r}+\frac{mp}{2G+1}}
    \quad{\rm and}\quad
    \bar{M}=\pmatrix{\frac{G+2}{r}+\frac{mp}{2G+1} &
    2\frac{\sqrt{G(G+1)}}{2G+1}mp \cr
    2\frac{\sqrt{G(G+1)}}{2G+1}mp &-\frac{G-1}{r}-\frac{mp}{2G+1}}
    \label{defM}
\end{equation}
where the Hankel function matrices $H_\pm$ are defined in
eq.~(\ref{hankel}).  The switch from lower to upper components is
given by
\begin{eqnarray}
    F_u&=&\frac{1}{\omega-ms}\sigma_3\left(F^\prime_l
    +F_uL_-^\prime+\bar{M}F_l\right) H_-H_+^{-1}
    \label{sw3}\\
    F_u^\prime&=&-F_uL_+^\prime-MF_l
    -(\omega+ms)\sigma_3F_uH_-H_+^{-1}\ .
    \label{sw4}
\end{eqnarray}
In the parity $-(-)^G$ channels the switch from upper to lower
components is given by
\begin{eqnarray}
    F_l&=&\frac{1}{\omega+ms}\sigma_3\left(F^\prime_u
    +F_uL_-^\prime+NF_u\right) H_-H_+^{-1}
    \label{sw5}\\
    F_l^\prime&=&-F_lL_+^\prime-\bar{N}F_l
    -(\omega-ms)\sigma_3F_uH_-H_+^{-1}
    \label{sw6}
\end{eqnarray}
with
\begin{equation}
    N=\pmatrix{\frac{G+2}{r}-\frac{mp}{2G+1} &
    2\frac{\sqrt{G(G+1)}}{2G+1}mp \cr
    2\frac{\sqrt{G(G+1)}}{2G+1}mp &-\frac{G-1}{r}+\frac{mp}{2G+1}}
    \quad{\rm and}\quad
    \bar{N}=\pmatrix{-\frac{G}{r}+\frac{mp}{2G+1} &
    2\frac{\sqrt{G(G+1)}}{2G+1}mp \cr
    2\frac{\sqrt{G(G+1)}}{2G+1}mp &\frac{G+1}{r}-\frac{mp}{2G+1}}
    \label{defN}
\end{equation}
and the transformation from lower to upper components is
\begin{eqnarray}
    F_u&=&\frac{-1}{\omega-ms}\sigma_3\left(F^\prime_l
    +F_uL_+^\prime+\bar{N}F_l\right) H_+H_-^{-1}
    \label{sw7}\\
    F_u^\prime&=&-F_uL_-^\prime-NF_l
    +(\omega+ms)\sigma_3F_uH_+H_-^{-1}\ .
    \label{sw8}
\end{eqnarray}
Finally, we note that the required ratios of Hankel functions can
also be expressed as rational functions
\begin{eqnarray}
    H_+H_-^{-1}&=&\frac{1}{k}\left[\sigma_3L_-^\prime+
    \frac{1}{r}\pmatrix{G+2 & 0\cr 0 & G-1}\right]
    \nonumber \\
    H_-H_+^{-1}&=&\frac{1}{k}\left[-\sigma_3L_-^\prime+
    \frac{1}{r}\pmatrix{G & 0\cr 0 & G+1}\right]\ .
    \label{Hrat}
\end{eqnarray}

\bigskip
\bigskip
\stepcounter{chapter}
\leftline{\large\bf Appendix B: Fake Boson Field}
\bigskip

In this Appendix, we describe the special treatment of the logarithmic
divergence of Feynman diagrams.  In particular we will explain and
provide numerical evidence for the use of the simplified form in
eqs.~(\ref{dbar}) and (\ref{ect4}).  
\bigskip
\bigskip

\leftline{\bf 1. Discussion}
\bigskip

In a fermion loop calculation, diagrams with one or two external
lines are quadratically divergent and those with three or four are
logarithmically divergent.  Hence we will have to take $N_{\rm max}=4$
in eq.~(\ref{ecasgeneral}).  Although we can straightforwardly compute
the corresponding Born terms as in eqs.~(\ref{Bfirst})
and~(\ref{Bsec}), the equivalent Feynman diagrams are difficult to
compute numerically\footnote{The fourth-order diagram requires a
nine-dimension integral, not including Fourier transforming the
background fields.} when the external fields are coordinate-dependent.
All we really need to do, however, is to regulate the momentum integral in
eq.~(\ref{ecasgeneral}) by subtracting an appropriate expression from
the integrand and ensure that we add back in exactly the same
quantity.  The latter quantity should have a divergent piece that
can easily be canceled by the counterterms.  Boson loops have a much
simpler divergence structure:  the logarithmic divergence corresponds
to a Feynman diagram that is only second order in the external lines.
Higher-order boson loop diagrams are finite.  We can therefore
simplify the regularization of the logarithmic divergence of the
fermion loop significantly by subtracting and adding back the
contributions of an equivalent boson.  This boson is completely
artificial to the model so we will call it the ``fake boson field.''
We impose the condition that at one loop the fake boson model
generates the same regulated logarithmic divergence as the original
fermion loop.  We then subtract the associated second-order  Born
phase shift from the fermion phase shifts and add it back in
as a regulated Feynman diagram. By construction, its
divergent piece is canceled by the counterterm contributions in
eq.~(\ref{ecasgeneral}).  This fake boson method does not give the
full Feynman diagrams of the fermion loop, so the approach is not
suitable to determine counterterm coefficients in a specified
renormalization prescription.  It can only be used once these
coefficients are known.  In our model we have uniquely determined the
counterterms from the first- and second-order terms in the expansion of
the fermion loop, which we computed exactly. We can then apply the
fake boson method to the third- and fourth-order terms, which would
otherwise be difficult to evaluate.  We can extract the local piece of
a Feynman diagram by setting the external momenta to zero.  An
expansion in the external momenta then shows that for a second-order
boson diagram only this local piece diverges.  We will identify the
local piece in the  second-order boson diagram as a ``limiting
function'' to the the second-order Born approximation to the phase
shift. This procedure provides the simplified expression used in
eqs.~(\ref{evac})--(\ref{ect4}).
\bigskip
\bigskip

\leftline{\bf 2. Equivalent Boson}
\bigskip

We begin by considering the second-order Born approximation to
a boson loop.  We consider the spherically symmetric problem
\begin{equation}
    -\frac{d^2}{dr^2}u_{\ell}(r)+
    \left[\frac{\ell(\ell+1)}{r^2}+gV(r)\right]u_{\ell}(r)
    =k^2u_{\ell}(r)
    \label{Brad1}
\end{equation}
discussed in Ref.~\cite{Fa98}.  The coupling constant $g$ is a
bookkeeping device which at the end we will take to be unity.  Solving
for the complex function $\beta_\ell(k,r)$ in the {\it ansatz}
$u_\ell(r)= {\rm exp}(2i\beta_\ell(k,r)) r h^{(1)}_\ell(kr)$ with the
boundary conditions
$\beta_\ell(k,\infty)=\beta^\prime_\ell(k,\infty)=0$ yields the phase
shifts \cite{Fa98}
\begin{equation}
    \delta_\ell(k)=-2{\rm Re} \beta_\ell(k,0)\ .
    \label{Bdelta}
\end{equation}
The differential equation for $\beta_\ell(k,r)$ is non-linear and
the expansion $\beta_\ell(k,r)=g \beta^{(1)}_\ell(k,r)+
g^2\beta^{(2)}_\ell(k,r) +\ldots$ yields the various orders of
the Born series to the phase shifts $\delta^{(n)}_\ell=-2{\rm Re}
\beta^{(n)}_\ell(k,0)$ by iteratively solving the differential
equations for $\beta^{(n)}_\ell(k,r)$.  $V(r)$ provides the source
term for $\beta^{(1)}_\ell(k,r)$.  Formally, the quadratically and
logarithmically divergent contributions to the vacuum polarization
energy are contained in
\begin{eqnarray}
    E^{(1)}_{\rm cas}&=&\int_0^\infty \frac{dk}{\pi}
    \sqrt{k^2+m^2}\frac{d}{dk}
    \sum_\ell(2\ell+1)\delta^{(1)}_\ell(k)
    \nonumber \\
    E^{(2)}_{\rm cas}&=&\int_0^\infty\frac{dk}{\pi}
    \sqrt{k^2+m^2}\frac{d}{dk}
    \sum_\ell(2\ell+1)\delta^{(2)}_\ell(k)\ .
    \label{Bcasexp}
\end{eqnarray}
Now let us consider two potentials, $V_1(r)$ and $V_2(r)$, which are
related by
\begin{equation}
    \int_0^\infty dr\,   r^2 V^2_1(r) = \int_0^\infty dr\,   r^2 V^2_2(r)
    \label{potcond}
\end{equation}
where we also allow for different masses $m_1$ and $m_2$ of the boson
field.  The dispersion relation $\omega=\sqrt{k^2+m_i^2}$ is
the only place where a dependence on the mass appears since
eq.~(\ref{Brad1}) does not contain the mass parameter explicitly.

Since the logarithmically divergent counterterm only depends on the
potential through the quantity $\int d^3 x V(x)^2$, it will be
identical for both potentials.  Therefore the difference of the
second-order Feynman diagrams
\begin{eqnarray}
    \Delta_{\rm F}&=&\frac{1}{(4\pi)^2}
    \int_0^\infty \frac{dq\,   q^2}{(2\pi)^2}
    \tilde{V_1}(q)\tilde{V_1}(-q)\left[-2+\sqrt{1+\frac{4m_1^2}{q^2}}\,
    \ln \frac{\sqrt{1+\frac{4m_1^2}{q^2}}+1}
    {\sqrt{1+\frac{4m_1^2}{q^2}}-1}\right]
    \nonumber \\ &&
    -\frac{1}{(4\pi)^2}\int_0^\infty \frac{dq\,   q^2}{(2\pi)^2}
    \tilde{V_2}(q)\tilde{V_2}(-q)\left[-2+\sqrt{1+\frac{4m_2^2}{q^2}}\,
    \ln \frac{\sqrt{1+\frac{4m_2^2}{q^2}}+1}
    {\sqrt{1+\frac{4m_2^2}{q^2}}-1}\right]
    \label{DeltaF}
\end{eqnarray}
will be finite.  Here $\tilde{V_i}(q)$ denotes the Fourier
transform of $V_i(r)$. Since we can identify orders in the Feynman
diagrams with those in the Born series for the vacuum polarization
energy, $\Delta_{\rm F}$ should be identical to
\begin{equation}
    \Delta_{\rm V}=E^{(2)}_{{\rm cas,}1}-E^{(2)}_{{\rm cas,}2}
    \label{DeltaV}
\end{equation}
where the difference is to be taken under the integral in the second
equation of~(\ref{Bcasexp}).  
The second subscript, $i=1,2$, refers to
    the potential $V_i$.
In Table~\ref{tab_B1} we compare
$\Delta_{\rm F}$ and $\Delta_{\rm V}$ for two Gaussian-type choices
\begin{equation}
    V_1(r)=d^2{\rm e}^{-r^2/w^2}+2d{\rm e}^{-r^2/2w^2}
    \qquad {\rm and} \qquad
    V_2(r)=C{\rm e}^{-r^2/2w^2}\ .
    \label{choices}
\end{equation}
Here $d$ and $w$ are variational parameters and the coefficient
$C$ is fixed by eq.~(\ref{potcond}).
\begin{table}
\caption{\label{tab_B1}Comparison of second-order contributions to
the vacuum polarization energy.}
\vskip0.2cm
\centerline{
\begin{tabular}{rccrr}
$d\,\,$ & $w$ & $m_2/m_1$ & $\Delta_V \,$\null & $\Delta_F\,$
\\ 
\hline
--2.0 & 2.0 & 1.0 & 0.088 & 0.087 \\
--2.0 & 2.0 & 0.5 & 6.915 & 6.916 \\
--2.0 & 2.0 & 2.0 &--7.309 &--7.312 \\
2.0 & 0.8 & 1.0 & 0.010 & 0.010 \\
2.0 & 1.5 & 1.0 & 0.023 & 0.023 \\
3.0 & 2.0 & 1.0 & 0.011 & 0.011
\end{tabular}}
\end{table}
We observe that the differences $\Delta_V$ and $\Delta_F$ agree within
the numerical accuracy even though either of them may be quite large
in magnitude, especially when the two masses are taken to be
different.  We conclude that we can subtract the second Born
approximation and add back in the second-order Feynman diagram
associated with bosonic fluctuations about $V_{1}$ to regulate the
logarithmic divergences encountered in the study of any other problem
with the same divergences.  The method we employ in the Yukawa model
is the generalization of this procedure to the case of fermions.  Note
that by considering the second-order Dirac equation ({\it cf.}\ Appendix~A)
the phase shift calculation in the Yukawa model is essentially that of a boson
field with derivative interactions.
\bigskip
\bigskip

\leftline{\bf 3. Limiting Function}
\bigskip

Next we extract a local contribution containing the logarithmic
divergence.  We will manipulate this expression so that it can be
substituted into the phase shift formula for the vacuum polarization
energy.  This procedure leads to further simplifications for evaluating
the fermion vacuum polarization energy.  Since these manipulations
involve divergent objects they are not rigorous results.  However, we
will be able to verify their validity numerically for a specific
background potentials.  This check is sufficient to justify the use of
these simplifications in the Yukawa model because we can always revert
to that specific potential using the arguments of the previous subsection.

We formally identify the local contribution by setting the external
momenta to zero in the second-order Feynman diagram
\begin{eqnarray}
    I_{\rm loc}&=&\frac{i}{2}\int d^3r\,  V^2(r)\int\frac{d^4k}{(2\pi)^4}
    \left[k^2-m^2-i\epsilon\right]^{-2}
    \nonumber \\
    &=&\frac{-1}{8}\int_0^\infty dr\,   r^2\, V^2(r)\int_0^\infty
    \frac{dk}{\pi}\sqrt{k^2+m^2}\frac{d}{dk}
    \left\{\frac{1}{m}\arctan \frac{m}{k}+
    \frac{k}{k^2+m^2}\right\}  \label{limfct1}
\end{eqnarray}
where we have carried out finite $k_0$ and angular integrals.
Nevertheless, these manipulations are formal since they involve the
logarithmically divergent $k$-integral.  However, so far we have only
manipulated the integrand.  It is worthwhile to note that the $k$
dependent function in the last integral equals that of the last term
in eq.~(\ref{dbar}).  We subtract the local contribution from the full
second-order Feynman diagram, giving
\begin{equation}
    \Delta E_{\rm F}^{(2)}=\frac{1}{(4\pi)^2}\int_0^\infty
    \frac{dq\,   q^2}{(2\pi)^2}\tilde{V}(q)\tilde{V}(-q)
    \left[-2+\sqrt{1+\frac{4m^2}{q^2}}\,
    \ln \frac{\sqrt{1+\frac{4m^2}{q^2}}+1}
    {\sqrt{1+\frac{4m^2}{q^2}}-1}\right]
    \label{Feynloc}
\end{equation}
which, of course, is finite for the potentials of the type
listed in eq.~(\ref{choices}). Similarly we can define a finite
second-order energy by subtracting the formal expression
from the second Born approximant to the vacuum polarization energy
\begin{equation}
    \Delta E_{\rm V}^{(2)}=\int_0^\infty \frac{dk}{\pi}
    \sqrt{k^2+m^2}\frac{d}{dk}
    \left[\left(\sum_\ell(2\ell+1)\delta^{(2)}(k)\right)
    -\delta^{(2)}_{\rm lim.~fct.}(k)\right]
    \label{Casloc}
    \end{equation}
where the limiting function is given by
    \begin{equation}
    \delta^{(2)}_{\rm lim.~fct.}(k)=-\frac{1}{8}
    \int_0^\infty dr\,   r^2 V^2(r)\left\{\frac{1}{m}
    \arctan \frac{m}{k}+\frac{k}{k^2+m^2}\right\}
    \label{limfct2}
\end{equation}
In Table~\ref{tab_B2} we compare $\Delta E_{\rm F}^{(2)}$ and $\Delta
E_{\rm V}^{(2)}$ for the background potential $V_1(r)$ given in
eq.~(\ref{choices}).  The particle in the loop of the local
contribution does not need to have the same mass as that in the full
Feynman diagram.
\begin{table}
\caption{\label{tab_B2}Comparison of the second-order
Feynman diagram including the local subtraction with the
corresponding expression for the vacuum polarization energy.}
\vskip0.2cm
\centerline{
\begin{tabular}{cccrr}
$d$ & $w$ & $m_2/m_1$ & $\Delta E_{\rm F}^{(2)}\!\!$ &
$\Delta E_{\rm V}^{(2)}\!\!$
\\ \hline
2.0 & 0.8 & 1.0 & 0.001 & 0.001 \\
2.0 & 1.5 & 1.0 & 0.002 & 0.002 \\
3.0 & 2.0 & 2.0 & 0.113 & 0.112 \\
1.0 & 1.5 & 0.8 & 0.069 & 0.069 \\
1.0 & 1.5 & 1.2 &--0.036 &--0.036 
\end{tabular}}
\end{table}
Within the numerical accuracy we do not find any differences.  

\bigskip
\bigskip
\stepcounter{chapter}
\leftline{\large\bf Appendix C: Derivative expansion}
\bigskip

Using the techniques developed in Ref.~\cite{Ai84} we compute the two
leading orders of the derivative expansion for the fermion determinant
\begin{equation}
    \Gamma(\Phi)=-i\,{\rm Tr}\, \ln \,\left\{i\dslash-g\Phi_5\right\}\ .
    \label{fdetC}
\end{equation}
First we compute the effective potential.  In dimensional regularization,
\begin{equation}
    V(\Phi)=\frac{2}{(4\pi)^2}\left[\frac{1}{\epsilon}-\gamma
    +\frac{3}{2}-\ln \,\frac{g^2\Phi^2}{4\pi\mu^2}\right]
    g^4\Phi^4
    \label{poteff}
\end{equation}
where $\mu$ is the scale introduced to render $g$ dimensionless in
$d=4-2\epsilon$ dimensions.  In order to extract the contribution with
two derivatives acting on $\Phi$ we parameterize
$\Phi=\Phi_0+\delta\Phi$ with $\Phi_0$ constant.  Since
\begin{equation}     \Gamma(\Phi_0+\delta\Phi)=-V(\Phi_0+\delta\Phi)
    +\frac{1}{2}\Gamma_{\mu\nu}^{(2)}(\Phi_0) \partial^\mu \delta\Phi
    \partial^\nu \delta\Phi +{\cal O}(\partial \delta\Phi)^4
    \label{d2general}
\end{equation}
it is sufficient to expand $\Gamma(\Phi_0+\delta\Phi)$ to quadratic
order in both $\delta\Phi$ and the momentum of its Fourier
transformation. Returning to configuration space this yields
\begin{eqnarray}
    &&\frac{i}{6}\int d^4x\int \frac{d^4q}{(2\pi)^4}\Bigg\{
    \frac{2}{(q^2-g^2\Phi_0^2)^2}
    {\rm tr}_F\left(\partial_\mu \delta\Phi \partial^\mu
    \delta\Phi^\dagger\right)
    \nonumber \\ && \hspace{1.5cm}
    +\frac{1}{(q^2-g^2\Phi_0^2)^3}
    \left[4q^2{\rm tr}_F\left(\partial_\mu \delta\Phi \partial^\mu
    \delta\Phi^\dagger\right)
    +4g^2{\rm tr}_F\left(\Phi_0^\dagger\partial_\mu \delta\Phi
    \Phi_0^\dagger\partial^\mu \delta\Phi\right)\right]\Bigg\}\ .
    \hspace{1.0cm}
    \label{d2first}
\end{eqnarray}
Now we replace $\Phi_0$ by $\Phi$ and $\partial_\mu\delta\Phi$ by
$\partial_\mu\Phi$ and treat the loop integral in dimensional
regularization,
\begin{eqnarray}
    &&\frac{g^2}{(4\pi)^2}\int d^4x
    \left(\frac{1}{\epsilon}-\gamma-\frac{3}{2}
    -\ln \,\frac{g^2\Phi^2}{4\pi\mu^2}\right)
    {\rm tr}_F\left(\partial_\mu \Phi \partial^\mu \Phi^\dagger\right)
    \nonumber \\ && \hspace{2.0cm}
    +\frac{1}{3}{\rm tr}_F\left(\partial_\mu \Phi \partial^\mu \Phi^\dagger
    -\frac{1}{|\Phi|^2}\Phi^\dagger\partial_\mu \Phi
    \Phi^\dagger\partial^\mu \Phi\right)\ .
    \label{d2second}
\end{eqnarray}
Combining the expressions eq.~(\ref{poteff}) and (\ref{d2second}) with
the counterterms computed in Section 3 yields the final result for the
derivative expansion up to second order,
\begin{eqnarray}
    {\cal L}^{(2)}&=&\frac{g^2}{(4\pi)^2}\Big\{
    \left[c_2-\ln \, \frac{|\Phi|^2}{v^2}\right]
    {\rm tr}_F\left(\partial_\mu \Phi \partial^\mu \Phi^\dagger\right)
    \nonumber \\ && \hspace{2.0cm}
    +\frac{1}{3}{\rm tr}_F
    \left(\partial_\mu \Phi \partial^\mu \Phi^\dagger
    -\frac{1}{|\Phi|^2}\Phi^\dagger\partial_\mu \Phi
    \Phi^\dagger\partial^\mu \Phi\right)\Big\}
    \nonumber \\ &&
    -\frac{g^4}{8\pi^2}(1+c_0)
    \left(|\Phi|^2-v^2\right)^2
    -\frac{g^4}{16\pi^2}\left[|\Phi|^4-v^4-
    2|\Phi|^4\ln \, \frac{|\Phi|^2}{v^2}\right]
    \label{gradres}
\end{eqnarray}
where
\begin{eqnarray}
    c_0&=&\frac{m_H^2}{4m^2}+\frac{3}{2}\int_0^1 dx\,
    \ln \,\left[1-x(1-x)\frac{m_H^2}{m^2}\right]
    \nonumber \\
    c_2&=&6\int_0^1 dx\, x(1-x)\,
    \ln \,\left[1-x(1-x)\frac{m_H^2}{m^2}\right]\ .
    \nonumber
\end{eqnarray}
We observe that for configurations with $|\Phi|^2>v^2$, the sum
of the classical energy and the contribution computed from
eq.~(\ref{gradres}) can become negative.  Such configurations will 
destabilize the vacuum.  Numerically, we have verified that this
behavior also emerges in the full calculation for the vacuum polarization
energy, which goes beyond the derivative expansion. Certainly this is an
artifact of the one-loop approximation, implying that within this
approximation we may not consider such configurations.

\end{document}